\documentclass[journal]{IEEEtran}

\usepackage{cite}      

\usepackage{graphicx}  

\usepackage{amsmath}   

\begin{document}
%
\title{Broadband SBS Slow Light in an Optical Fiber}
%
%
\author{Zhaoming~Zhu,~\IEEEmembership{Member,~OSA,}
        Andrew~M.~C.~Dawes,~\IEEEmembership{Student Member,~OSA,}
        Daniel~J.~Gauthier,~\IEEEmembership{Fellow,~OSA,}
        Lin~Zhang,~\IEEEmembership{Student Member,~IEEE, Student
        Member,~OSA}, and~Alan~E.~Willner,~\IEEEmembership{Fellow,~IEEE, Fellow,~OSA}%
\thanks{This work was supported by DARPA DSO Slow-Light program.}
\thanks{Z.~Zhu, A.M.C. Dawes and D.J. Gauthier are with the Department of Physics and
the Fitzpatrick Center for Photonics and Communications Systems,
Duke University, Durham, NC 27708, USA.}%
\thanks{L. Zhang and A.E. Willner are with the Department of Electrical and
Computer Engineering, University of Southern California, Los
Angeles, CA 90089, USA.}}

%
\markboth{Journal of Lightwave Technology}{Zhu
\MakeLowercase{\textit{et al.}}: Broadband SBS Slow Light in an
Optical Fiber}
%



\maketitle

\begin{abstract}
We investigate slow-light via stimulated Brillouin scattering in a
room temperature optical fiber that is pumped by a spectrally
broadened laser.  Broadening the spectrum of the pump field
increases the linewidth $\Delta\omega_p$ of the Stokes amplifying
resonance, thereby increasing the slow-light bandwidth.  One
physical bandwidth limitation occurs when the linewidth becomes
several times larger than the Brillouin frequency shift $\Omega_B$
so that the anti-Stokes absorbing resonance cancels out
substantially the Stokes amplifying resonance and hence the
slow-light effect.  We find that partial overlap of the Stokes and
anti-Stokes resonances can actually lead to an enhancement of the
slow-light delay - bandwidth product when $\Delta\omega_p \simeq 1.3
\Omega_B$.  Using this general approach, we increase the Brillouin
slow-light bandwidth to over 12 GHz from its nominal linewidth of
$\sim$30 MHz obtained for monochromatic pumping. We controllably
delay 75-ps-long pulses by up to 47 ps and study the data pattern
dependence of the broadband SBS slow-light system.
\end{abstract}

\begin{keywords}
Slow Light, Stimulated Brillouin Scattering, Optical Fiber, Pulse
Propagation, Q penalty.
\end{keywords}

%
\IEEEpeerreviewmaketitle

\section{Introduction}
\PARstart{T}{here} has been great interest in slowing the
propagation speed of optical pulses (so-called slow light) using
coherent optical methods~\cite{Gauthier_Boyd}. Slow-light techniques
have many applications for future optical communication systems,
including optical buffering, data synchronization, optical memories,
and signal processing~\cite{Gauthier_PhysicsWorld_2005,
Gauthier_PhotonicsSpectra_2006}. It is usually achieved with
resonant effects that cause large normal dispersion in a narrow
spectral region (approximately equal to the resonance width), which
increases the group index and thus reduces the group velocity of
optical pulses. Optical resonances associated with stimulated
Brillouin scattering
(SBS)~\cite{Okawachi_PRL_2005}--\cite{Zhu_JOSAB_2005}, stimulated
Raman scattering~\cite{Sharping_OE_2005} and parametric
amplification~\cite{Dahan_OE_2005} in optical fibers have been used
recently to achieve slow light.

The width of the resonance enabling the slow-light effect limits the
minimum duration of the optical pulse that can be effectively
delayed without much distortion, and therefore limits the maximum
data rate of the optical system~\cite{Stenner_2005}. In this regard,
fiber-based SBS slow light is limited to data rates less than a few
tens of Mb/s due to the narrow Brillouin resonance width ($\sim$30
MHz in standard single-mode optical fibers). Recently, Herr\'{a}ez
\emph{et al}.~\cite{Herraez_OE_2006} increased the SBS slow-light
bandwidth to about 325 MHz by broadening the spectrum of the SBS
pump field. Here, we investigate the fundamental limitations of this
method and extend their work to achieve a SBS slow-light bandwidth
as large as 12.6 GHz, thereby supporting data rates of over 10 Gb/s
\cite{Zhu_OFC_2006}. With our setup, we delay 75-ps pulses by up to
47 ps and study the data pulse quality degradation in the broadband
slow-light system.

This paper is organized as follows. The next section describes the
broadband-pump method for increasing the SBS slow-light bandwidth
and discuss its limitations.  Section \ref{sect3} presents the
experimental results of broadband SBS slow light, where we
investigate the delay of single and multiple pulses passing through
the system.  From the multiple-pulse data, we estimate the
degradation of the eye-diagram as a function of delay, a first step
toward understanding performance penalties incurred by this
slow-light method.  Section \ref{sect4} concludes the paper.

\section{SBS Slow Light}

In a SBS slow-light system, a continuous-wave (CW) laser beam
(angular frequency $\omega_p$) propagates through an optical fiber,
which we take as the $-z$-direction, giving rise to amplifying and
absorbing resonances due to the process of electrostriction. A
counterpropagating beam (along the $+z$-direction) experiences
amplification in the vicinity of the Stokes frequency
$\omega_s=\omega_p-\Omega_B$, where $\Omega_B$ is the Brillouin
frequency shift, and absorption in the vicinity of the anti-Stokes
frequency $\omega_{as}=\omega_p+\Omega_B$.

A pulse (denoted interchangeably by the ``probe'' or ``data'' pulse)
launched along the $+z$-direction experiences slow (fast) light
propagation when its carrier frequency $\omega$ is set to the
amplifying (absorbing)
resonance~\cite{Okawachi_PRL_2005}--\cite{Zhu_JOSAB_2005}. In the
small-signal regime, the output pulse spectrum is related to the
input spectrum through the relation
$E(z=L,\omega)=E(z=0,\omega)\exp[g(\omega)L/2]$, where $L$ is the
fiber length and $g(\omega)$ is the complex SBS gain function. The
complex gain function is the convolution of the intrinsic SBS gain
spectrum $\tilde{g}_{0}(\omega)$ and the power spectrum of the pump
field $I_p(\omega_p)$ and is given by
\begin{align}\label{Eq:convolution}
g(\omega) &= \tilde{g}_0(\omega) \otimes I_p(\omega_p)\\ \nonumber
&=\int_{-\infty}^{\infty} \frac{g_0 I_p(\omega_p)}{1-i(\omega +
\Omega_B -\omega_p)/(\Gamma_B/2)} d\omega_p,
\end{align}
where $g_0$ is linecenter SBS gain coefficient for a monochromatic
pump field, and $\Gamma_B$ is the intrinsic SBS resonance linewidth
(FWHM in radians/s). The real (imaginary) part of $g(\omega)$ is
related to the gain (refractive index) profile arising from the SBS
resonance.

In the case of a monochromatic pump field, $I_p(\omega_p)=I_0
\delta(\omega_p - \omega_{p0})$, and hence $g(\omega)=g_0 I_0
/[1-i(\omega+\Omega_B -\omega_{p0})/(\Gamma_B/2)]$; the gain profile
is Lorentzian. For a data pulse whose duration is much longer than
the Brillouin lifetime $1/\Gamma_B$ tuned to the Stokes resonance
($\omega=\omega_s$), the SBS slow-light delay is given by
$T_{del}=G_0/\Gamma_B$ where $G_0=g_0 I_0 L$ is the gain parameter
and $\exp(G_0)$ is the small-signal
gain~\cite{Okawachi_PRL_2005}--\cite{Zhu_JOSAB_2005}.  The SBS
slow-light bandwidth is given approximately by $\Gamma_B/2\pi$ (FWHM
in cycles/s).

Equation (\ref{Eq:convolution}) shows that the width of the SBS
amplifying resonance can be increased by using a broadband pump.
Regardless of the shape of the pump power spectrum, the resultant
SBS spectrum is approximately equal to the pump spectrum when the
pump bandwidth is much larger than the intrinsic SBS linewidth.
This increased bandwidth comes at some expense:  the SBS gain
coefficient scales inversely with the bandwidth, which must be
compensated using a higher pump intensity or using a fiber with
larger $g_0$.

To develop a quantitative model of the broadband SBS slow-light, we
consider a pump source with a Gaussian power spectrum, as realized
in our experiment. To simplify the analysis, we first consider the
case when the width of the pump-spectrum broadened Stokes and
anti-Stokes resonances is small in comparison to $\Omega_B$, which
is the condition of the experiment of Ref.~\cite{Herraez_OE_2006}.
Later, we will relax this assumption and consider the case when
$\Delta\omega_p\sim\Omega_B$ where the two resonances begin to
overlap, which is the case of our experiment.

In our analysis, we take the pump power spectrum as
\begin{equation}
I_p(\omega_p)=\frac{I_0}{\sqrt{\pi}\Delta\omega_p}
\exp\left[-\left(\frac{\omega_p-\omega_{p0}}{\Delta
\omega_p}\right)^2 \right ]. \label{Eq:pump-spec}
\end{equation}
Inserting this expression into Eq.~(\ref{Eq:convolution}) and
evaluating the integral results in a complex SBS gain function given
by
\begin{equation}
g(\omega)=g_0 I_0 \sqrt{\pi}\eta \text{w}(\xi+i\eta),
\label{Eq:gain-profile}
\end{equation}
where $\text{w}(\xi+i\eta)$ is the complex error
function~\cite{Handbook}, $\xi = (\omega+\Omega_B
-\omega_{p0})/\Delta\omega_p$, and $\eta
=\Gamma_B/(2\Delta\omega_p)$.

When $\eta \ll 1$ (the condition of our experiment), the gain
function is given approximately by
\begin{equation}
g(\omega)=g_0 I_0 \sqrt{\pi}\eta \exp(-\xi^2)\text{erfc}(-i\xi),
\label{Eq:gain-profile-approx}
\end{equation}
where erfc is the complementary error function.  The width (FWHM,
rad/s) of the gain profile is given by
$\Gamma=2\sqrt{\text{ln}~2}\Delta\omega_p$, which should be compared
to the unbroadened resonance width $\Gamma_B$. The line-center gain
of the broadened resonance is given by $G=\sqrt{\pi}\eta G_0$.

The SBS slow-light delay at line center for the broadened resonance
is given by
\begin{equation}
T_{del}=\frac{d {\rm
Im}[g(\omega)L/2]}{d\omega}|_{\omega=\omega_{s}} =
\frac{2\sqrt{\text{ln}~2}}{\sqrt{\pi}}\frac{G}{\Gamma} \approx
0.94\frac{G}{\Gamma}. \label{Eq:delay}
\end{equation}
A Gaussian pulse of initial pulse width $T_0$ ($1/e$ intensity
half-width) exits the medium with a broader pulse width $T_{out}$
determined through the relation
\begin{equation}
T_{out}^2=T_0^2+\frac{G}{\Delta\omega_p^2}. \label{Eq:pulse-width}
\end{equation}
Assuming that a slow-light application can tolerate no more than a
factor of two increase in the input pulse width ($T_{out}=2T_0$),
the maximum attainable delay is given by
\begin{equation}
\left(\frac{T_{del}^{max}}{T_o}\right)=\frac{3}{\sqrt{\pi}}T_0
\Delta\omega_p, \label{Eq:max-delay}
\end{equation}
which is somewhat greater than that found for a Lorentzian
line~\cite{Boyd_2005}.  From Eq.~(\ref{Eq:max-delay}), it is seen
that large absolute delays for fixed $\Delta\omega_p$ can be
obtained by taking $T_0$ large.

\begin{figure}
\centering
\includegraphics[width=0.45\textwidth]{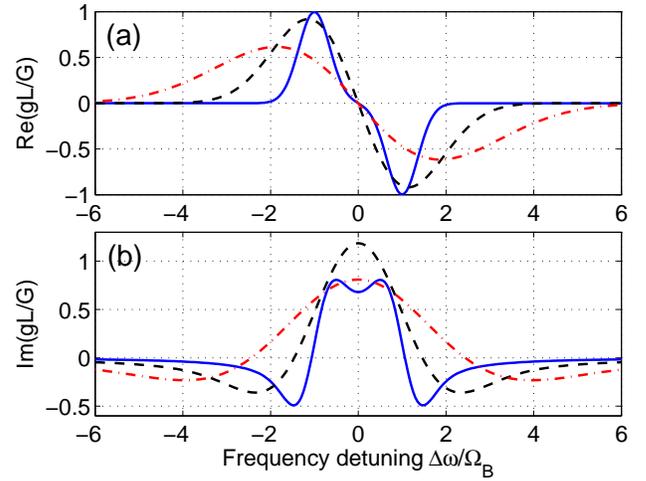}
\caption{SBS gain profiles at different pump power spectrum
bandwidth $\Delta \omega_p$: (a) real part and (b) imaginary part of
$g(\omega)$ as a function of frequency detuning from the pump
frequency. Solid curves: $\Delta \omega_p/\Omega_B=0.5$, dashed
curves: $\Delta \omega_p/\Omega_B=1.3$, dashed-dotted curves:
$\Delta \omega_p/\Omega_B=2.5$.} \label{Fig:gain-profiles}
\end{figure}

\begin{figure}[htb]
\centering
\includegraphics[width=0.45\textwidth]{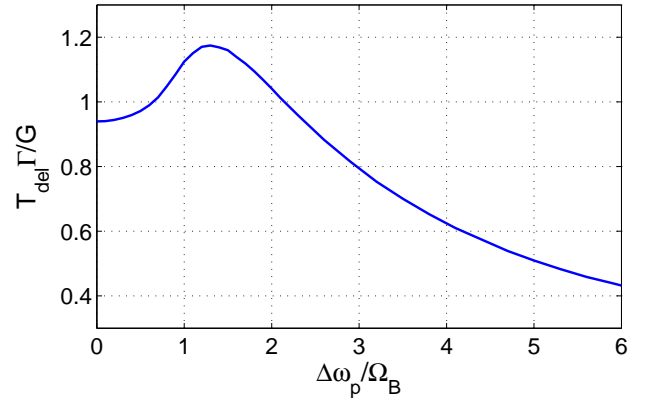}
\caption{Relative SBS delay as a function of the SBS resonance
linewidth.} \label{Fig:optimum-linewidth}
\end{figure}

We now turn to the case when the pump spectral bandwidth $\Delta
\omega_p$ is comparable with the Brillouin shift $\Omega_B$. In this
situation, the gain feature at the Stokes frequency
$\omega_{p0}-\Omega_B$ overlaps with the absorption feature at the
anti-Stokes frequency $\omega_{p0}+\Omega_B$.  The combination of
both features results in a complex gain function given by
\begin{equation}
g(\omega)=\frac{G}{L} \left({\rm e}^{-\xi_+^2}\text{erfc}(-i\xi_+)-
{\rm e}^{-\xi_-^2}\text{erfc}(-i\xi_-)\right),
\end{equation}
where $\xi_{\pm}=(\omega \pm \Omega_B -\omega_{p0})/\Delta\omega_p$.
As shown in Fig.~\ref{Fig:gain-profiles}, the anti-Stokes absorption
shifts the effective peak of the SBS gain to lower frequencies when
$\Delta \omega_p$ is large, and reduces the slope of the linear
phase-shift region and hence the slow-light delay. For intermediate
values of $\Delta\omega_p$, slow-light delay arising from the wings
of the anti-Stokes resonances enhances the delay at the center of
the Stokes resonance.  Therefore, there is an optimum value of the
resonance linewidth that maximizes the delay.
Figure~\ref{Fig:optimum-linewidth} shows the relative delay as a
function of the resonance bandwidth, where it is seen that the
optimum value occurs at $\Delta\omega_p \sim$ 1.3 $\Omega_B$ and
that the delay falls off only slowly for large resonance bandwidths.
This result demonstrates that it is possible to obtain practical
slow-light bandwidths that can somewhat exceed a few times
$\Omega_B$.

\section{Experiments and Results}\label{sect3}

As discussed above, the SBS slow-light pulse delay $T_{del}$ is
proportional to $G/\Gamma$. The decrease in $G$ that accompanies the
increase in $\Delta\omega_p$ needs to be compensated by increasing
the fiber length, pump power, and/or using highly nonlinear optical
fibers (HNLF). In our experiment, we use a 2-km-long HNLF (OFS,
Denmark) that has a smaller effective modal area and therefore a
larger SBS gain coefficient $g_0$ when compared with a standard
single-mode optical fiber. We also use a high-power Erbium-doped
fiber amplifier (EDFA, IPG Model EAD-1K-C) to provide enough pump
power to achieve appreciable gain.

\begin{figure}[tb]
\centering
\includegraphics[width=0.48\textwidth]{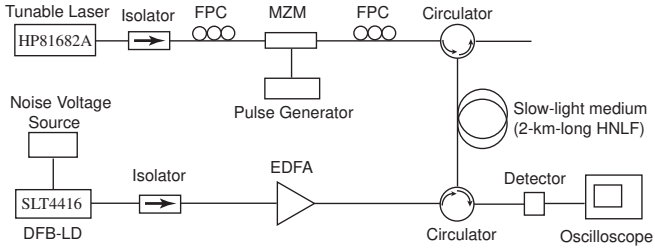}
\caption{Experiment setup. EDFA: Erbium-doped fiber amplifier, MZM:
Mach-Zehnder modulator, FPC: fiber polarization controller, HNLF:
highly nonlinear fiber. } \label{Fig:setup}
\end{figure}

To achieve a broadband pump source, we directly modulate the
injection current of a distributed feedback (DFB) single-mode
semiconductor laser. The change in injection current changes the
refractive index of the laser gain medium and thus the laser
frequency, which is proportional to the current-modulation
amplitude. We use an arbitrary waveform generator (TEK, AWG2040) to
create a Gaussian noise source at a 400-MHz clock frequency, which
is amplified and summed with the DC injection current of a 1550-nm
DFB laser diode (Sumitomo Electric, STL4416) via a bias-T with an
input impedance of 50 Ohms. The resultant laser power spectrum is
approximately Gaussian. The pump power spectral bandwidth is
adjusted by changing the peak-peak voltage of the noise source.

The experiment setup is shown schematically in Fig.~\ref{Fig:setup}.
Broadband laser light from the noise-current-modulated DFB laser
diode is amplified by the EDFA and enters the HNLF via a circulator.
The Brillouin frequency shift of the HNLF is measured to be
$\Omega_B/2\pi$ = 9.6 GHz. CW light from another tunable laser is
amplitude-modulated to form data pulses that counter-propagate in
the HNLF with respect to the pump wave. Two fiber polarization
controllers (FPC) are used to maximize the transmission through the
intensity modulator and the SBS gain in the slow-light medium. The
amplified and delayed data pulses are routed out of the system via a
circulator and detected by a fast photoreceiver (12-GHz bandwidth,
New Focus Model 1544B) and displayed on a 50-GHz-bandwidth sampling
oscilloscope (Agilent 86100A). The pulse delay is determined from
the waveform traces displayed on the oscilloscope.

\begin{figure} [bth]
\centering
\includegraphics[width=0.48\textwidth]{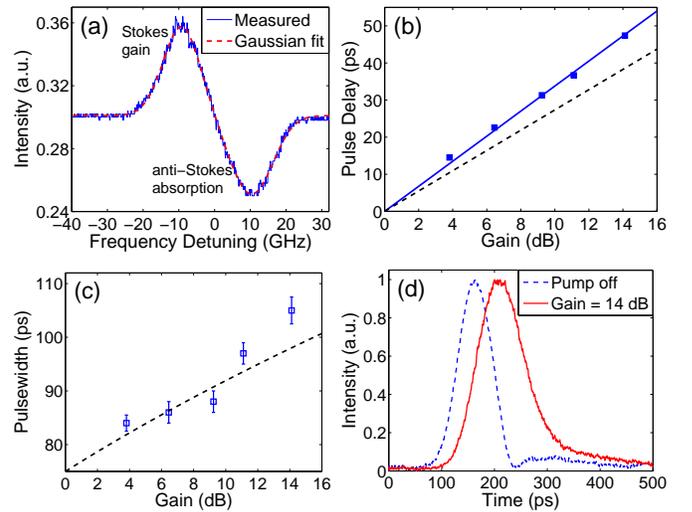}
\caption{Observation of broadband slow-light delay. (a) Measured SBS
gain spectrum with a dual Gaussian fit. The SBS gain bandwidth
(FWHM) is found to be 12.6 GHz. Pulse delay (b) and pulse width (c)
as a function of SBS gain. In (b), the solid line is the linear fit
of the measured data (solid squares), and the dashed line is
obtained with Eq.~(\ref{Eq:delay}). In (c), the dashed curve is
obtained with Eq.~(\ref{Eq:pulse-width}). (d) Pulse waveforms at
0-dB and 14-dB SBS gain. The input data pulsewidth is $\sim$75 ps.}
\label{Fig:delay-data}
\end{figure}

To quantify the effect of the bandwidth-broadened pump laser on the
SBS process, we measured the broadened SBS gain spectra by scanning
the wavelength of a CW laser beam and measuring the resultant
transmission. Figure~\ref{Fig:delay-data}(a) shows an example of the
spectra.  It is seen that the features overlap and that
Eq.~(\ref{Eq:gain-profile-approx}) does an excellent job in
predicting our observations, where we adjusted $\Gamma$ to obtain
the best fit.  We find $\Gamma/2\pi$ = 12.6 GHz
($\Delta\omega_p/\Omega_B\sim 0.8$), which is somewhat smaller than
the optimum value.  We did not attempt to investigate higher
bandwidths to avoid overdriving the laser with the broadband signal.
This non-ideality could be avoided by using a laser with a greater
tuning sensitivity.

Based on the measured SBS bandwidth, we chose a pulsewidth (FWHM) of
$\sim$75 ps ($T_0 \sim$ 45 ps) produced by a 14 Gb/s electrical
pulse generator. Figures \ref{Fig:delay-data}(b)-(d) show the
experimental results for such input pulses. Figure
\ref{Fig:delay-data}(b) shows the pulse delay as a function of the
gain experienced by the pulse, which is determined by measuring the
change in the pulse height. A 47-ps SBS slow-light delay is achieved
at a pump power of $\sim$580 mW that is coupled into the HNLF, which
gives a gain of about 14 dB. It is seen that the pulse delay scales
linearly with the gain, demonstrating the ability to control
all-optically the slow-light delay. The dashed line in
Fig.~\ref{Fig:delay-data}(b) is obtained with Eq.~(\ref{Eq:delay}),
which tends to underestimate the time delay that is enhanced by the
contribution from the anti-Stokes line (see
Fig.~\ref{Fig:optimum-linewidth}). Figure \ref{Fig:delay-data}(c)
shows the width of the delayed pulse as a function of gain. The data
pulse is seen to be broadened as it is delayed, where it is
broadened by about 40\% at a delay of about 47 ps. The dashed curve
in Fig.~\ref{Fig:delay-data}(c) is obtained with
Eq.~(\ref{Eq:pulse-width}). Figure \ref{Fig:delay-data}(d) shows the
waveforms of the undelayed and delayed pulses at a gain of 14 dB. We
observe pulse delays that are due to fiber lengthening under strong
pump conditions due to fiber heating. These thermally-induced delays
are not included in Fig.~\ref{Fig:delay-data}(b).

\begin{figure}[htb]
\centering
\includegraphics[width=0.48\textwidth]{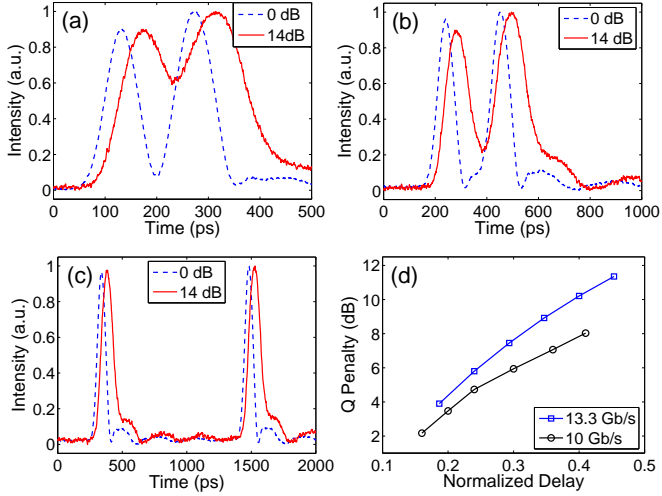}
\caption{Pattern dependence of SBS slow-light delay. (a) Data pulses
of pattern `101.' (b) Data pulses of pattern `1001.' Note the change
in the horizontal scale. (c) Data pulse of pattern
`10000000000000001.' In (a)-(c), the data bit-rate is 14 Gb/s and
the input single pulsewidth is $\sim$75 ps. (d) Calculated Q penalty
vs. normalized time delay for 13.3 Gb/s and 10 Gb/s bit-rate data. }
\label{Fig:pattern}
\end{figure}

To investigate how the pulse broadening seen in
Fig.~\ref{Fig:delay-data}(c) might impact a communication system, we
examine the pattern dependence of the pulse distortion.  For
example, in NRZ data format, a single `1' pulse has a different gain
than consecutive `1' pulses~\cite{Zhang_OFC_06}. The
pattern-dependent gain could induce a different `1' level in the
whole data stream, while pattern-dependent delay can lead to a large
timing jitter.

Figures \ref{Fig:pattern}(a)-(c) show the delayed pulse waveforms of
three simple NRZ data patterns with a bit-rate of 14 Gb/s. It is
clear that the pulses overlap when they are closer to each other,
which degrades the system performance. To quantify the signal
quality degradation, we use Q-factor (signal quality factor) of
input and output pulses, which is defined as
$(m_1-m_0)/(\sigma_1+\sigma_0)$, where $m_1$, $m_0$, $\sigma_1$,
$\sigma_0$ are the mean and standard deviation of the signal samples
when a `1' or `0' is received. We examine the Q-penalty (decrease in
Q-factor) produced by the broadband SBS slow-light system by
numerical simulations. Figure \ref{Fig:pattern}(d) shows the
Q-penalty as a function of time delay for 10 Gb/s and 13.3 Gb/s
bit-rate data streams, respectively. In the simulations, the `1'
pulse is assumed to be Gaussian-shaped with a pulsewidth (FWHM) of
the bit time (100 ps for 10 Gb/s, 75 ps for 13.3 Gb/s). The
slow-light delay is normalized by the bit time so that Q-penalties
in different bit-rate systems can be compared. It is seen that the
Q-penalty increases approximately linearly with the normalized
delay, and that the 13.3 Gb/s data rate incurs a higher penalty than
the 10 Gb/s data rate. The penalty is higher at the higher data rate
because the higher-speed signal is more vulnerable to the pattern
dependence, especially when the slow-light bandwidth is comparable
to the signal bandwidth. Error-free transmission (BER $<10^{-9}$) is
found at a normalized delay of 0.25 or less. In an optimized system,
it is expected that the pattern dependence can be decreased using a
spectrum-efficient signal modulation format or the signal carrier
frequency detuning technique~\cite{Zhang_OFC_06}, for example.

\section{Conclusion}\label{sect4}
In summary, we have increased the bandwidth of SBS slow light in an
optical fiber to over 12 GHz by spectrally broadening the pump
laser, thus demonstrating that it can be integrated into existing
data systems operating over 10 Gb/s. We observed a pattern
dependence whose power penalty increases with increasing slow-light
delay; research is underway to decrease this dependence and improve
the performance of the high-bandwidth SBS slow-light system.

\section*{Acknowledgment}
We gratefully acknowledge the loan of the fast pulse generator and
sampling oscilloscope by Martin Brooke of the Duke Electrical and
Computer Engineering Department.

\begin{biographynophoto}{Zhaoming Zhu}
received a Bachelor degree in Electronic Engineering and an
M.S. degree in Applied Physics from Tsinghua University, Beijing,
China, in 1995 and 1998, respectively, and a Ph.D. degree in Optics
from the University of Rochester in 2004. His Ph.D. research on
``Photonic crystal fibers: characterization and supercontinuum
generation" was supervised by Prof. T.~G. Brown. Currently, he is a
postdoctoral research associate under the mentorship of Prof. D.~J.
Gauthier at Duke University studying optical-fiber-based slow light
effects and applications. His research interests include nonlinear
optics, guided-wave and fiber optics, and photonic crystals.

Dr. Zhu is a member of the Optical Society of America and the
American Physical Society.
\end{biographynophoto}

\begin{biography}[{\includegraphics[width=1in,height=1.25in,clip,keepaspectratio]{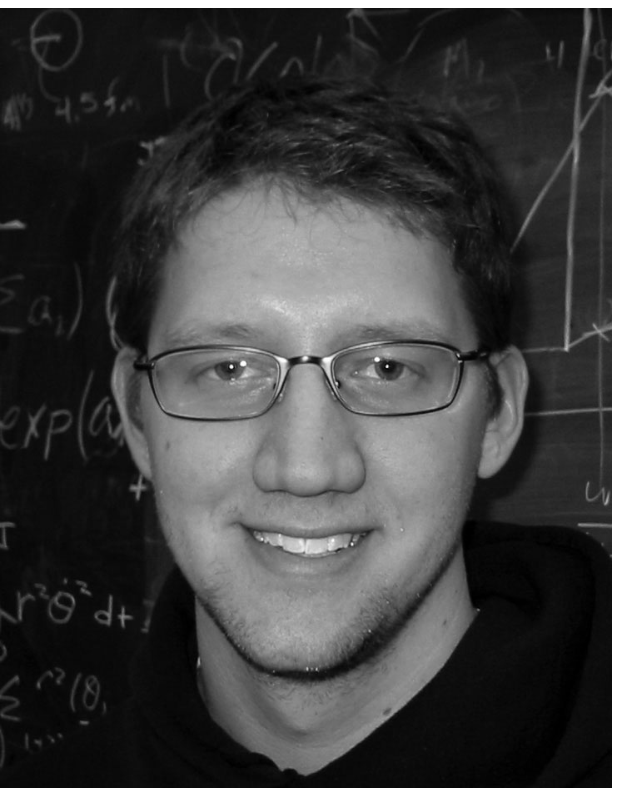}}]{Andrew M. C. Dawes}
received the B.A. degree with honors in physics from Whitman
College, Walla Walla, WA, and the M.A. degree in physics from Duke
University, Durham, NC in 2002 and 2005 respectively. He is
currently pursuing the Ph.D. degree in the Duke University
Department of Physics. His research interests include slow-light in
optical fiber, pattern formation in nonlinear optics, and
all-optical switching and processing systems. Mr. Dawes is a student
member of the Optical Society of America (OSA) and the American
Physical Society (APS) and currently a Walter Gordy Graduate Fellow
of the Duke University Department of Physics and a John T. Chambers
Fellow of the Fitzpatrick Center for Photonics and Communications
Systems.
\end{biography}

\begin{biography}[{\includegraphics[width=1in,height=1.25in,clip,keepaspectratio]{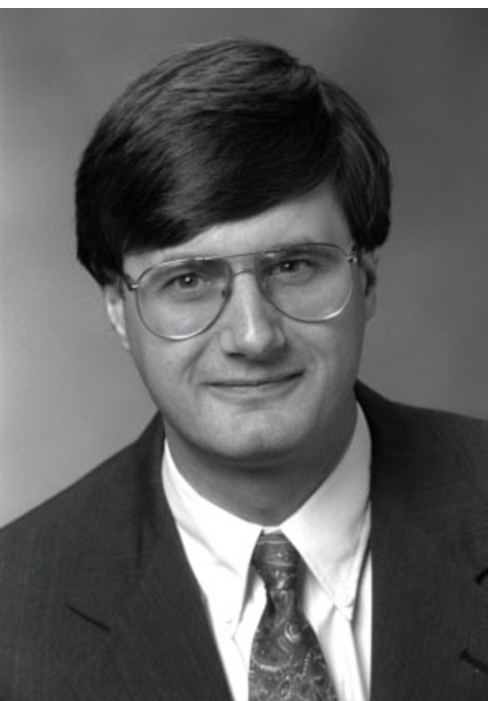}}]{Daniel J. Gauthier}
received the B.S., M.S., and Ph.D. degrees from the University of
Rochester, Rochester, NY, in 1982, 1983, and 1989, respectively. His
Ph.D. research on ``Instabilities and chaos of laser beams
propagating through nonlinear optical media" was supervised by Prof.
R.~W. Boyd and supported in part through a University Research
Initiative Fellowship.

From 1989 to 1991, he developed the first CW two-photon optical
laser as a Post-Doctoral Research Associate under the mentorship of
Prof. T.~W. Mossberg at the University of Oregon. In 1991, he joined
the faculty of Duke University, Durham, NC, as an Assistant
Professor of Physics and was named a Young Investigator of the U.S.
Army Research Office in 1992 and the National Science Foundation in
1993.

He is currently the Anne T. and Robert M. Bass Professor of Physics
and Biomedical Engineering at Duke. His research interests include:
applications of slow light in classical and quantum information
processing and controlling and synchronizing the dynamics of complex
electronic, optical, and biological systems.

Prof. Gauthier is a Fellow of the Optical Society of America and the
American Physical Society.
\end{biography}

\begin{biography}[{\includegraphics[width=1in,height=1.25in,clip,keepaspectratio]{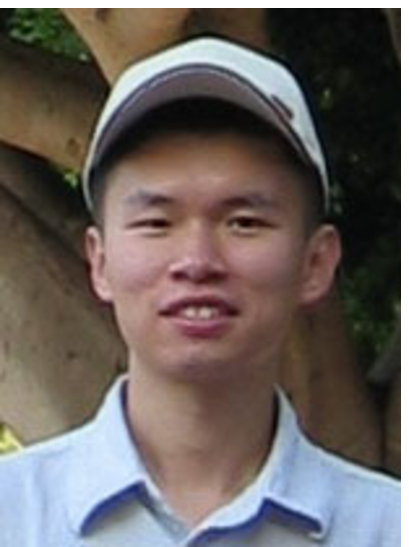}}]{Lin Zhang}
was born in Anshan, Liaoning, China, in 1978. He received the B.S.
and M.S. degree from Tsinghua University, Beijing, China, in 2001
and 2004, respectively. His thesis was on birefringence and
polarization dependent coupling in photonic crystal fibers. Now he
is pursuing the Ph.D. degree in the Department of Electrical
Engineering, the University of Southern California, Los Angeles. His
current research interests include fiber-based slow light, photonic
crystal fibers, nonlinear optics, and fiber optical communication
systems.

Lin Zhang is a student member of the Optical Society America (OSA)
and IEEE Lasers and Electro-Optics Society (LEOS). He was awarded as
one of top-ten outstanding graduate students of 2003 year at
Tsinghua University.
\end{biography}

\begin{biography}[{\includegraphics[width=1in,height=1.25in,clip,keepaspectratio]{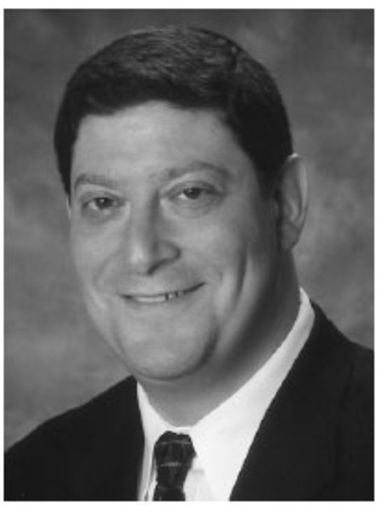}}]{Alan E. Willner}
(S'87-M'88-SM'93-F'04) received the Ph.D. degree from Columbia
University, New York. He has worked at AT\&T Bell Laboratories and
Bellcore. He is currently Professor of Electrical Engineering at the
University of Southern California (USC), Los Angeles. He has 525
publications, including one book.

Prof. Willner is a Fellow of the Optical Society of America (OSA)
and was a Fellow of the Semiconductor Research Corporation. He has
received the NSF Presidential Faculty Fellows Award from the White
House, the Packard Foundation Fellowship, the NSF National Young
Investigator Award, the Fulbright Foundation Senior Scholars Award,
the IEEE Lasers \& Electro-Optics Society (LEOS) Distinguished
Traveling Lecturer Award, the USC University-Wide Award for
Excellence in Teaching, the Eddy Award from Pennwell for the Best
Contributed Technical Article, and the Armstrong Foundation Memorial
Prize.

His professional activities have included: President of IEEE LEOS,
Editor-in-Chief of the IEEE/OSA JOURNAL OF LIGHTWAVE TECHNOLOGY,
Editor-in-Chief of the IEEE JOURNAL OF SELECTED TOPICS IN QUANTUM
ELECTRONICS, Co-Chair of the OSA Science and Engineering Council,
General Co-Chair of the Conference on Lasers and Electro-Optics
(CLEO), General Chair of the LEOS Annual Meeting Program, Program
Co-Chair of the OSA Annual Meeting, and Steering and Program
Committee Member of the Conference on Optical Fiber Communications
(OFC).
\end{biography}

\end{document}